\documentclass[twocolumn]{aastex62} 
 
\usepackage{graphicx}
\usepackage{natbib,aas_macros}
\received{January 26, 2019} 
\revised{March 4, 2019} 
\accepted{March 22, 2019} 
 
\shorttitle{Outer Perseus Spiral Arm}                                                               
\shortauthors{Sakai et al.} 
 
\begin{document} 
 
\title{Non-circular Motions in the Outer Perseus Spiral Arm}
 

\correspondingauthor{Nobuyuki Sakai} 
\email{nobuyuki.sakai@nao.ac.jp}
\email{nsakai@kasi.re.kr}
 
\author[0000-0002-0786-7307]{Nobuyuki Sakai} 
\affil{National Astronomical Observatory of Japan, 2-21-1 Osawa,
Mitaka, Tokyo 181-8588, Japan}                                          
 \affil{Korea Astronomy $\&$ Space Science Institute, 776, Daedeokdae-ro, 
Yuseong-gu, Daejeon 34055, Korea}                                          

\author{Mark J. Reid} 
\affiliation{Center for Astrophysics $\vert$ Harvard \& Smithsonian, 
60 Garden Street, Cambridge, MA 02138, USA}                  
 
 
\author{Karl M. Menten} 
\affiliation{Max-Planck-Institut f{\"u}r Radioastronomie Auf dem
H{\"u}gel 69, D-53121 Bonn, Germany}                                    
 
\author{Andreas Brunthaler} 
\affiliation{Max-Planck-Institut f{\"u}r Radioastronomie Auf dem
H{\"u}gel 69, D-53121 Bonn, Germany}                                    
 
\author{Thomas M. Dame} 
\affiliation{Center for Astrophysics $\vert$ Harvard \& Smithsonian, 
60 Garden Street, Cambridge, MA 02138, USA}

\begin{abstract} 
 
We report measurements of parallax and proper motion for five 6.7-GHz 
methanol maser sources in the outer regions of the Perseus arm as part of the BeSSeL
Survey of the Galaxy. By combining our results with previous astrometric
results, we determine an average spiral arm pitch angle of $9.2\pm1.5$ deg 
and an arm width of 0.39 kpc for this spiral arm.   
For sources in the interior side of the Perseus arm, 
we find on average a radial inward motion in the Galaxy of $13.3\pm5.4$ km s$^{-1}$ 
and counter to Galactic rotation of $6.2\pm3.2$ km s$^{-1}$. 
These characteristics are consistent with models for spiral arm formation that involve 
gas entering an arm to be shocked and then forming stars.  However, similar data for
other spiral arms do not show similar characteristics. 
 
\end{abstract} 
 


\section{Introduction} \label{sec:intro} 

While spiral patterns can be prominent in disk galaxies, their
formation mechanism and the dynamical evolution of spiral arms remain
under discussion.  Two general mechanisms for spiral arm formation have
dominated the discussion in the literature: 
1) density-wave theories \citep{1964ApJ...140..646L, 1969ApJ...158..123R} and 
2) dynamic theories \citep{1965MNRAS.130..125G, 1969ApJ...158..899T}.  
In density-wave theories, spiral structures are long-lived and rotate 
nearly uniformly, while stars and gas rotate differentially and pass through 
the arms.
In dynamic theories, arms are short-lived and reform as open structures.
They are seen in $N$-body simulations of multi-arm spirals (e.g.,
\citealp{1984ApJ...282...61S, 2000Ap&SS.272...31S, 2010arXiv1001.5430S, 2002MNRAS.336..785S, 2005A&A...444....1F, 2011ApJ...730..109F}),
unbarred grand-design spirals (\citealp{2011MNRAS.410.1637S}), and barred spirals (e.g.,
\citealp{2009ApJ...706..471B, 2012MNRAS.426..167G, 2013MNRAS.432.2878R, 2015MNRAS.454.2954B}).                                  
 
Comparing the spatial distributions of stars and gas in a spiral arm
may help to distinguish between the two mechanisms 
(e.g., \citealp{2014PASA...31...35D, 2015PASJ...67...69S, 2018ApJ...853L..23B, 2017MNRAS.465..460E}).
Density-wave theories predict a 
spatial offset between the gravitational potential minimum of a spiral 
arm (traced by the distribution of old stars) and the peak of gas density 
(\citealp{1969ApJ...158..123R}). On the other hand, dynamic (non-stationary) spiral-arm models
predict that both star and gas accumulate into a minimum in the
spiral potential and hence are not separated 
(e.g., \citealp{2008MNRAS.385.1893D, 2011ApJ...735....1W, 2015PASJ...67L...4B}). 
In the future one could test these theories by comparing the 3-dimensional (3-d) positions of 
older stars measured by {\it Gaia} with that of masers from newly 
formed stars measured by Very Long Baseline Interferometry (VLBI).

If gas entering a spiral arm is shocked prior to the formation of stars, 
the resulting stars should display the kinematic signature of that shock.  
By measuring 3-d velocity fields, one could therefore determine if such 
shocks occur and how strong they are. Such as the 
Bar and Spiral Structure Legacy (BeSSeL) Survey and VLBI Exploration of Radio Astrometry
(VERA) have yielded precise distances and 3-d velocity fields for
high-mass star-forming regions (HMSFRs) associated with spiral arms 
(e.g., \citealp{2009ApJ...700..137R, 2012PASJ...64..136H, 2014ApJ...783..130R}). 
Optical astrometric results from $Gaia$ DR2 typically have parallax uncertainties 
larger than 20 $\mu$as (e.g. see Fig. 7 in \citealp{2018A&A...616A...1G}) and are starting 
to become significant for this type of study.

Recently, \citet{2006Sci...311...54X}, \citet{2012PASJ...64..108S} and \citet{2014ApJ...790...99C} showed evidence for 
systematic (radially) inward motion for HMSFRs in the Perseus arm.         
Here, we report new astrometric results obtained with the National Radio Astronomy
Observatory (NRAO)\footnote{NRAO official HP: \\ \url{https://public.nrao.edu/}}
 Very Long Baseline Array (VLBA), which more clearly 
reveal the structure and kinematics of the Perseus arm. 
In section 2, we describe our VLBA observations.
In section 3, we outline the data reduction. In section 4,
we show new astrometric results for five 6.7-GHz CH$_{3}$OH masers. 
In section 5, we discuss the structure
and kinematics of the Perseus arm, based on our new results and {\it Gaia}
DR2 results for OB-type stars (taken from \citealp{2018A&A...616L..15X}) and compare
those with spiral arm models. In section 6, we summarize the paper.

\section{Observation} \label{sec:obs} 
 
We observed a total of six methanol masers 
(ie, the CH$_{3}$OH ($J_{K}$ = 5$_{1} -$ 6$_{0}A^{+}$) transition
at a rest frequency of 6.668519 GHz) under VLBA programs
BR149S, T and U (see Table \ref{table:4} in the appendix)\footnote{Please
also see the BeSSeL survey HP: \\ \url{http://bessel.vlbi-astrometry.org/observations}}.  
Each set of observations was optimized to sample the peaks of the sinusoidal 
parallax signature in right ascension over one year, as described for previous
BeSSeL Survey observations (e.g., \citealp{2016SciA....2E0878X,2017AJ....154...63R}). 
Each maser source, listed in Table \ref{table:4}, was observed with three or 
four background quasars (QSOs).  A single observation involved  
(i) four half-hour ``geodetic blocks'' spaced by about 2 hours for clock and atmospheric 
delay calibration, (ii) ``manual phase-calibration'' scans of a bright quasar (QSO) every 2 hours,
iii) fast switching between a target maser and each QSO, used for relative position determination.
 
Observational data was recorded on the Mark5A system at 512 Mbps. 
Geodetic block data was taken in left circular polarization with
four 16 MHz bands spanning 496 MHz centered at both 4.3 and 7.3 GHz
(8 IFs in total).   Fast switching data was taken
in dual circular polarization with four adjacent 16-MHz bands spanning
64 MHz. The data were correlated with the DiFX software correlator
\citep{2011PASP..123..275D} in Socorro, NM.                                           
The fast switching data were correlated in two passes:
for the maser (line) data the central
8 MHz of the third IF band was correlated with 1000 channels, giving
a frequency (velocity) spacing of 8 kHz (0.36 km s$^{-1}$) at the
rest frequency.  The continuum data for all IFs were correlated
with 32 spectral channels.

\section{Data reduction} \label{sec:analysis} 

The VLBA data reduction was conducted with the NRAO Astronomical Image
Processing System (AIPS) and a ParselTongue pipeline described in
previous BeSSeL Survey papers (e.g., \citealp{2009ApJ...693..397R}). Details of
the techniques employed to determine parallax and proper motion for
6.7-GHz CH$_{3}$OH masers are described in \citet{2017AJ....154...63R}. Here, we
briefly outline the data reduction.
 
The largest source of relative position error for 6.7 GHz astrometric data is 
uncertainty in the ionospheric delay calibration. For the
ionospheric delay calibration, we firstly applied the Global Ionospheric
Maps obtained from NASA's ftp server\footnote{\url{ftp://cddis.gsfc.nasa.gov/gps/products/ionex/}}.
However, at our observing frequency of 6.7 GHz, tropospheric and ionospheric
delay residuals can still be significant, with residual path-delays of
$\sim$5 cm for both components.   Using the geodetic-block observations,
tropospheric (non-dispersive) delays were estimated by differencing delays 
at 4.3 and 7.3 GHz and subtracting these from the total delays.  These
were modeled as owing to a zenith delay for each observations block.
To better calibrate the ionospheric delay residual, the delay differences
between 4.3 and 7.3 GHz bands were scaled to the 6.7 GHz CH$_{3}$OH band
and a residual zenith dispersive delay could also be determined. 
 
After applying the geodetic-block calibrations to the phase-reference
data, we used a bright maser spot as the phase reference for the 
associated QSOs.  In cases where the maser displayed significant
structure, we self-calibrated the maser data and applied these
solutions to both maser and QSO data.  All sources were imaged
and the positions of compact components were determined by fitting
elliptical Gaussian brightness distributions.   The variations the
positions of maser spots relative to background QSOs were then modeled
as owing to parallax and proper motion components.

Delay residuals at 6.7 GHz are generally dominated by the ionospheric 
miscalibration and can cause a systematic position shift across the sky 
(so called ionospheric wedges).  We used our multiple QSO data to
account for these effects.  As discussed in \citet{2017AJ....154...63R}, one can
generate an ``artificial QSO'' at the position of the target maser
to remove most of the effects of the ionospheric wedges.  In this
paper we incorporated an improved procedure that solved for the
wedge effects at each epoch, while at the same time estimating the
parallax and proper motion components as described in \citet{2019arXiv190109313W}.

\section{Results} \label{sec:results} 
 
\begin{figure*}[tbhp] 
 \begin{center} 
     \includegraphics[scale=1.12]{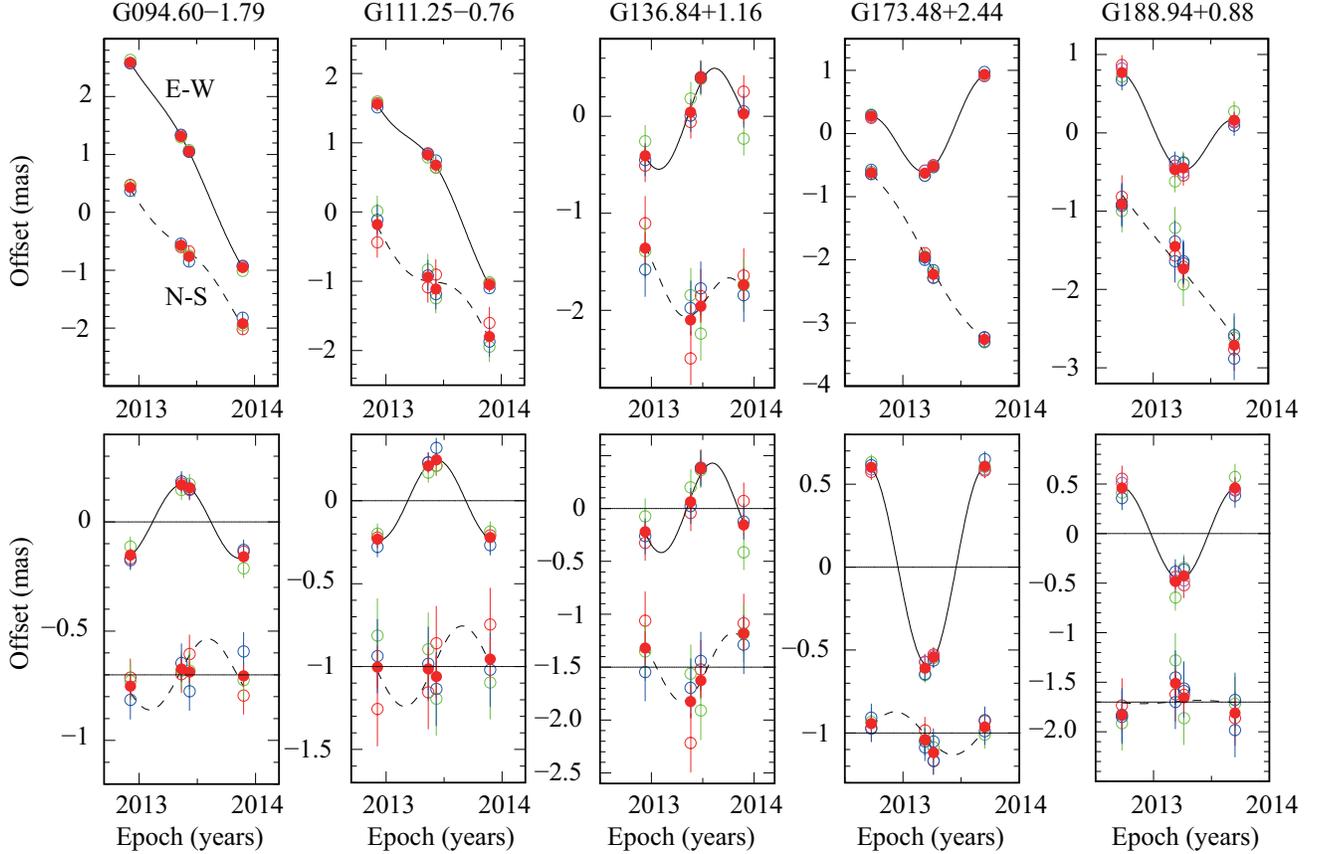} 
\end{center} 
\caption{ Results of parallax and proper-motion 
fitting. Plotted are position offsets of maser spots with respect to background
QSOs toward the east (RAcos$\delta$) and north ($\delta$) 
as a function of time.  For clarity, the northerly data is plotted offset from the
easterly data. Colored open-circles show the position
offsets relative to 1st (red), 2nd (blue), 3rd (green) and 4th QSOs
(purple) (see Table \ref{table:4} in the Appendix). Filled
circles represent unweighted averages of the QSO results. 
Error bars of the average QSO results are too small to be seen in most
cases. \textit{(Top row)} The best-fit models in the easterly and northerly directions are shown
as continuous and dashed curves, respectively.  
\textit{(Bottom row)} Same as top row, but with proper motions removed.}
\label{fig:1} 
\end{figure*}  

We estimated trigonometric parallaxes and proper motions for five of the
six sources we observed. For G098.03+1.44 
the brightest maser spot was too faint ($<0.5$ Jy) to use as a phase reference.
In order to estimate a single proper motion for each source, we 
averaged the values for all spots.  We adopted these values for
the proper motion of the central star, but added $\pm5$ km s$^{-1}$
in quadrature to the fitted error estimates for each motion coordinate in 
order to allow for uncertainty in this step.   Note that class II CH$_{3}$OH masers 
generally have internal motions of about 5 km s$^{-1}$ (e.g., \citealp{2010ApJ...716.1356M}).
The parallax and proper motions results are summarized in Table \ref{table:1} (see also Fig. \ref{fig:1}).

Some sources have published parallax and proper 
motions for 22-GHz water or 12.2-GHz methanol masers as discussed below.   
When combining parallaxes, we used variance weighting.   However, for
combined proper motion estimates of 6.7-GHz methanol and 22-GHz water masers motions, 
we adopted the methanol values, since water masers typically form in outflows 
of tens of km s$^{-1}$, and transferring the maser motions to that of the central star 
is less certain than from methanol masers.

We now briefly discuss some individual sources.

\subsection{G094.60$-$1.79}
 
All background QSOs were northward of the maser, rather than
surrounding it.  When accounting for the effects of ionospheric wedges, this
requires extrapolation of the fitted planar position tilt instead of interpolation,
and we expect increased parallax uncertainty.  To allow for this, 
we added $\pm0.05$ mas per degree of offset times the offset of the nearest QSO in
quadrature with the formal parallax uncertainty.
This parallax gradient error source is a rough estimate based on
BeSSeL Survey experience fitting for 6.7 GHz data for many sources.  

\citet{2010PASJ...62..101O} used the VERA array and estimated a parallax of 
$0.326\pm0.031$ mas based on three 22-GHz water maser spots.  
In order to be conservative, we have inflated their uncertainty by
$\sqrt{3}$ to $\pm0.054$ mas in order to allow for correlated systematic errors caused 
by similar differential atmospheric delay differences between maser spots and 
a background QSO. Generally, residual atmospheric delay errors dominate cm-wave VLBI parallax uncertainty. Another result for this source comes from \citet{2014ApJ...790...99C}, 
who obtained a 22-GHz parallax of $0.253\pm0.024$ mas using the VLBA.
In order to assess if the three parallax results are statistically consistent 
we calculated parallax differences of $0.147\pm0.072$, $0.074\pm0.054$, and $0.073\pm0.059$ mas.
Only the first difference is marginally statistically significant, while the other two 
are statistically insignificant.  We conclude that these could reasonably have come from random
differences and combine all three by variance-weighting to give a best parallax for 
G094.60$-$1.79 of $0.250\pm0.020$ mas.
 
\subsection{G111.25$-$0.76} 

\citet{2014ApJ...790...99C} derived a parallax of $0.299\pm0.022$ mas for water masers and
the difference between this and our result is not statistically significant ($0.035\pm0.030$ mas). 
Thus we variance-weighted them to obtain a best parallax of 0.280$\pm$0.015 mas.

\subsection{G136.84+1.16} 

All background QSOs were northward of the maser, and, as discussed above for G094.60$-$1.70, 
we inflated the parallax uncertainty to account for a likely $\pm0.05$ (mas/degree) parallax 
gradient.  However, since the maser's structure was fairly extended, the final parallax 
uncertainty is quite large ($\pm0.123$ mas).

\subsection{G188.94+0.88} 

\citet{2010PASJ...62..101O} using VERA obtained a parallax of $0.569\pm0.068$ mas for water masers, 
while \citet{2009ApJ...693..397R} using the VLBA found a value of $0.476\pm0.006$ mas based on 
12-GHz methanol masers.  The differences among the three parallax measurements
are not statistically significant ($0.104\pm0.080$, $0.011\pm0.042$, and
$0.093\pm0.068$ mas), and we variance weighted them to obtain a best parallax of
0.476$\pm$0.006 mas. 
For a combined proper motion, we use our 6.7-GHz and the published
12-GHz methanol maser result, yielding ($\mu_{\alpha}\rm{cos}\delta$,
$\mu_{\delta}$) = ($-0.30\pm0.60$, $-1.95\pm0.51$) mas yr$^{-1}$,
where we have added in quadrature $\pm5$ km s$^{-1}$ for each component uncertainty.

\begin{table*}[htbp] 
\caption{6.7 GHz parallax and proper motion results} 
\label{table:1} 
\begin{center} 
\begin{tabular}{lcrr} 
\tableline 
\tableline 
Source	&Parallax  &$\mu_{\alpha}$cos$\delta \hspace{1em}	$&$\mu_{\delta}
\hspace{2em}$	\\                                                        
             &(mas)&(mas yr$^{-1}$)	&(mas yr$^{-1}$)	\\ 
\tableline 
G094.60$-$1.79  &0.179$\pm$0.048   &$-$3.49$\pm$0.22   &$-$2.65$\pm$0.23 \\
G111.25$-$0.76  &0.264$\pm$0.020   &$-$2.69$\pm$0.28   &$-$1.75$\pm$0.33 \\
G136.84+1.16    &0.442$\pm$0.123   &0.43$\pm$0.48      &$-$0.45$\pm$0.56 \\
G173.48+2.44    &0.594$\pm$0.014   &0.62$\pm$0.63      &$-$2.34$\pm$0.63 \\
G188.94+0.88    &0.465$\pm$0.042   &$-$0.62$\pm$0.51   &$-$1.87$\pm$0.53 \\
\tableline 
\end{tabular} 
\end{center} 
\tablecomments{Column 1 gives the source name, and column 2 gives our 6.7-GHz
parallax result.   Columns 3 and 4 list our measured  proper-motion results
in the eastward and northward directions, respectively. These motions are
meant to represent those of the central star which excites the masers.}     
           
\end{table*} 
 
\clearpage

\section{Discussion} 
\begin{figure*}[tbhp] 
 \begin{center} 
     \includegraphics[scale=2.5]{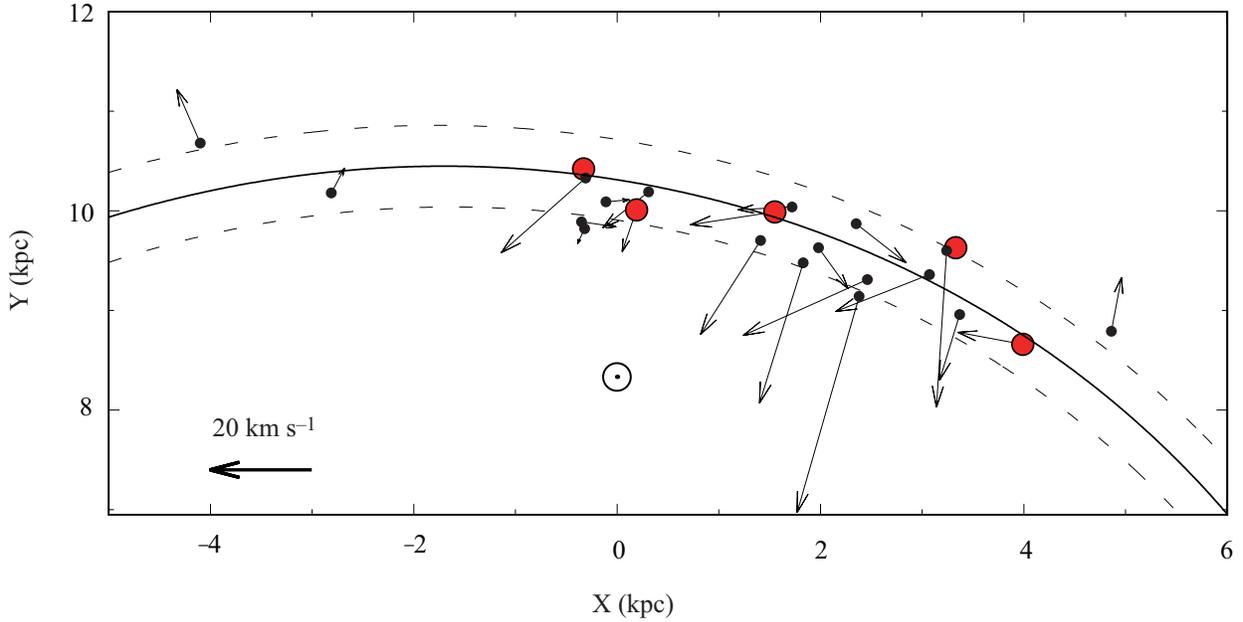} 
\end{center} 
\caption{Spatial distribution and non-circular motions of Perseus-arm
sources. Large red and small black
filled circles show our and previous VLBI astrometric results.
For clarity, only sources whose motion uncertainties are less than 20 km s$^{-1}$ are plotted.
A scale of 20 km s$^{-1}$ (thick arrow) is displayed at the lower
left.  The solid curve represents a fitted logarithmic spiral 
model and the dashed lines indicate $\pm1\sigma$ width. 
The Sun is at (X,Y) = (0, 8.34) kpc.  The non-circular motions are with respect to
Galactic constants ($R_{0}=8.34$ kpc and $\Theta_{0}=241$ km s$^{-1}$) and a solar motion 
of ($U_{\odot}$, $V_{\odot}$, $W_{\odot}$) = (10.5, 14.4, 8.9) km s$^{-1}$ \citep{2014ApJ...783..130R}.}                                                           
\label{fig:2} 
\end{figure*} 

\subsection{Pitch angle and arm width of the Perseus arm} \label{sec:5.1}  
Using the five astrometric results discussed above, along with 
other sources in the literature, we now evaluate the pitch angle and width 
of the Perseus arm based on 27 sources.       
Following \citet{2014ApJ...783..130R}, we fitted a logarithmic spiral-arm model to the 
locations of the Perseus arm sources using the following equation:                       
 
\begin{equation} 
\mathrm{ln}(R/R_{\rm{ref}})=-(\beta-\beta_{\rm{ref}})\mathrm{tan}\psi, 
\end{equation}

\noindent  
where $R_{{\rm ref}}$ and $\beta_{{\rm ref}}$ are a Galactocentric
radius (kpc) at a reference azimuth (radians).  Azimuth ($\beta$) is defined as zero
toward the Sun as viewed from the Galactic center and increases with
Galactic longitude, and $\psi$ is the spiral pitch angle.  $\beta_{{\rm ref}}$ = 13.2 degrees
was chosen to be  near the midpoint of the azimuth values.
Our best fitting values are $R_{{\rm ref}}$ =  9.93 $\pm$ 0.09 kpc and 
$\psi$ = 9.2 $\pm$ 1.5 degrees.   These results are consistent with those in 
\citet{2014ApJ...783..130R} within errors.  The arm's width, defined 
as the $1\sigma$ scatter in the sources perpendicular to the
fitted arm, is 0.39 kpc.

\subsection{Non-circular motion in the Perseus arm} \label{sec:5.2}

We now assess the three-dimensional non-circular (peculiar) motions of Perseus arm sources, 
based on the parallaxes, proper motions and LSR velocities. The LSR velocity of the central star is estimated 
from the masers and from observations of thermal line emission (e.g., CO) from the parent cloud. Peculiar motions are 
referenced to a model Galactic rotation curve, using Galactic parameters ($R_{0}$ and $\Theta_{0}$), 
and solar motion ($U_{\odot}$, $V_{\odot}$, $W_{\odot}$). 
Note that $U$ values are positive directed toward the Galactic center, $V$ is in the 
direction of the Galactic rotation, and $W$ is toward the north Galactic pole.                    
In the following, we use 24 of the 27 sources available, removing three outliers
(G043.16+0.01, owing to its large motion uncertainty of $>20$ km s$^{-1}$, and 
G108.20+0.58 and G229.57+0.15 owing to their large deviation of $>$2.2$\sigma$ 
from the spiral arm fit).

\begin{figure*}[tbhp] 
 \begin{center} 
     \includegraphics[scale=1.0]{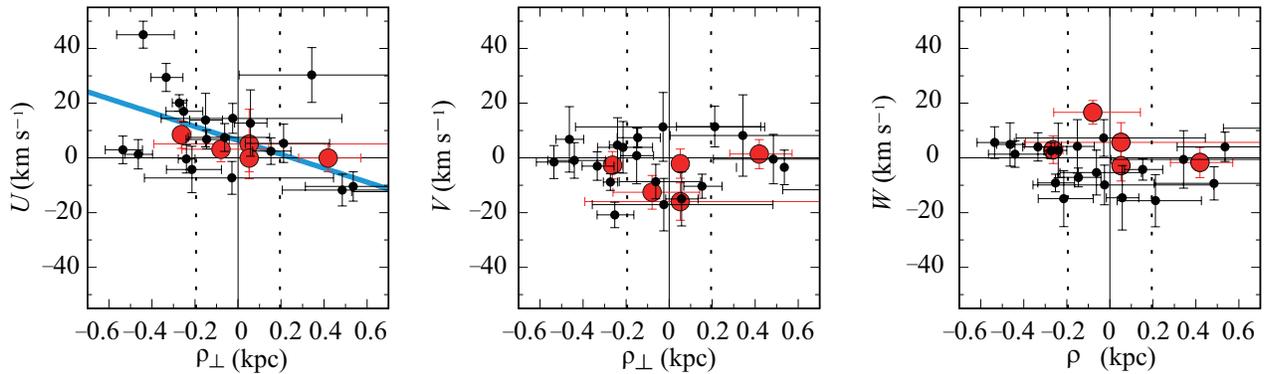} 
\end{center} 
    \caption{{\bf(Left)} Non-circular motion toward the Galactic center
($U$) as a function of distance perpendicular to the center of the Perseus arm 
($\rho_{\perp}$) for VLBI astrometric results.
Large red and small black filled circles show our and previous VLBI astrometric
 results, respectively. Note that positive $\rho_{\perp}$ value means
 exterior of the Perseus arm as viewed from the Galactic center. The vertical dashed
 line represents $\pm$0.5$\sigma_{\rm{Per}}$ (kpc), where $\sigma_{\rm{Per}}= 0.39$ kpc. 
The cyan line highlights a significant slope ($>3\sigma$)
for a weighted least-squares fit giving $U=-$25.4($\pm$8.2)$\rho_{\perp}$+6.4($\pm$2.4). 
{\bf(Middle)} Same as (Left), but for non-circular motion in the direction of the Galactic
rotation ($V$). No significant velocity gradient was found for this component.
{\bf (Right)} Same as (Left), but for non-circular motion toward
the north Galactic pole ($W$). As for $W$, no significant velocity gradient was obtained.}
    \label{fig:3} 
\end{figure*}

\begin{table*}[htbp]
\caption{Statistics for non-circular motion across the Perseus arm.}
\label{table:2}
\begin{center} 
\begin{tabular}{ccrrcrrrc}
\tableline
\tableline
	\multicolumn{3}{c}{Interior side}&\multicolumn{3}{c}{Middle region}&\multicolumn{3}{c}{Exterior}\\
	\multicolumn{3}{c}{$-\frac{3\sigma_{\rm{Per}}}{2} \leqq  \rho_{\perp} \rm{(kpc)} < -\frac{\sigma_{\rm{Per}}}{2}$}&\multicolumn{3}{c}{$-\frac{\sigma_{\rm{Per}}}{2} \leqq \rho_{\perp} \rm{(kpc)} < \frac{\sigma_{\rm{Per}}}{2}$}&\multicolumn{3}{c}{$\frac{\sigma}{2} \leqq  \rho_{\perp} \rm{(kpc)} < \frac{3\sigma}{2}$}\\
\cline{1-3} \cline{4-6} \cline{7-9}
$<$$U$$>$ &$<$$V$$>$ 	&\colhead{$<$$W$$>$} 	& $<$$U$$>$ &$<$$V$$>$&\colhead{$<$$W$$>$} & $<$$U$$>$  &$<$$V$$>$ &\colhead{$<$$W$$>$}\\
	(km s$^{-1}$)	&(km s$^{-1})$ &(km s$^{-1}$)	&(km s$^{-1}$)&(km s$^{-1}$)&(km s$^{-1}$) &(km s$^{-1}$)&(km s$^{-1}$)&(km s$^{-1}$)\\
\cline{1-3} \cline{4-6} \cline{7-9}
\multicolumn{3}{c}{(9 masers)} &\multicolumn{3}{c}{(10 masers)} &\multicolumn{3}{c}{(5 masers)}\\
{\bf 13.3$\pm$5.4}		&$-$2.5$\pm$2.8	&0.0$\pm$2.4		&{\bf 5.9$\pm$2.1}	&$-$6.2$\pm$3.2	&$-$1.0$\pm$3.0	&2.7$\pm$7.6	&3.4$\pm$2.8&$-$4.6$\pm$3.5	\\
\tableline
\end{tabular}
\end{center}
\tablecomments{Columns 1-3 represent unweighted means of the non-circular motion components ($U$, $V$, $W$) for masers at
  $-\frac{3\sigma_{\rm{Per}}}{2} \leqq  \rho_{\perp} \rm{(kpc)} < -\frac{\sigma_{\rm{Per}}}{2}$, where $\sigma_{\rm{Per}}$ (= 0.39 kpc)
  is the arm width of the Perseus arm.  The uncertainties are the standard error of the mean. The 
numbers of sources available are indicated in parentheses.
A number with the bold font emphasizes a statistical significance greater than $2\sigma$.
Columns 4-6 are  for masers at $-\frac{\sigma_{\rm{Per}}}{2} \leqq \rho_{\perp} \rm{(kpc)} < \frac{\sigma_{\rm{Per}}}{2}$.
  Columns 7-9 are for masers at $\frac{\sigma_{\rm{Per}}}{2} \leqq  \rho_{\perp} \rm{(kpc)} < \frac{3\sigma_{\rm{Per}}}{2}$.}
\end{table*}

Figure \ref{fig:2} shows the non-circular motions of Perseus-arm
sources with uncertainties less than 20 km s$^{-1}$. 
The solid curve in Fig. \ref{fig:2} represents a logarithmic 
spiral-arm fit and the dashed lines indicate ($\pm1\sigma$) width (see section \ref{sec:5.1}). 
Figure \ref{fig:3} plots the non-circular motions as a function of distance perpendicular 
to the arm, $\rho_{\perp}$, defined positive outward from the arm.
Interestingly, we see a significant velocity gradient of $-25\pm8$ km s$^{-1}$ kpc$^{-1}$ 
in $U$ vs. $\rho_{\perp}$.  Note that excluding the two high points with $U>25$ km s$^{-1}$ 
and $\rho_{\perp}<-0.2$ kpc as potential outliers still yields a significant gradient of 
$-16\pm7$ km s$^{-1}$ kpc$^{-1}$.  The other peculiar motion components ($V,W$) do not show a 
statistically significant gradient across the spiral arm.
 
As an alternative approach to examining systematics in the peculiar motions,
Table \ref{table:2} presents unweighted means of ($<$$U$$>$, $<$$V$$>$, $<$$W$$>$) 
in three $\rho_{\perp}$ bins:  
the interior given by 
$-\frac{3\sigma_{\rm{Per}}}{2} \leqq  \rho_{\perp} \rm{(kpc)} < -\frac{\sigma_{\rm{Per}}}{2}$),
the middle given by 
($-\frac{\sigma_{\rm{Per}}}{2}$ $\leqq \rho_{\perp}$ (kpc) $< \frac{\sigma_{\rm{Per}}}{2}$),
and the exterior given by 
($\frac{\sigma_{\rm{Per}}}{2} \leqq  \rho_{\perp} \rm{(kpc)} < \frac{3\sigma_{\rm{Per}}}{2}$).
In the above, $\sigma_{\rm{Per}}$ is a Gaussian $1\sigma$ width for the arm, which
we estimate to be 0.39 kpc.  For uncertainties, we adopt the standard error of the mean, 
because the scatter evident in Figures \ref{fig:3}(Left) is much larger than 
would be suggested by the measurement uncertainties, indicating there is significant 
``astrophysical'' noise. 

As anticipated by the negative gradient of $U$ vs $\rho_{\perp}<0$,
the results in Table \ref{table:2} show that sources 
toward the interior side of the arm are moving radially inward with 
$<$$U$$>$ $= 13.3\pm5.4$ km s$^{-1}$ (2.5$\sigma$ for 9 masers), while
sources exterior to the arm show a small average $U$ motion of
$=2.7\pm7.6$ km s$^{-1}$ (for 5 masers). 
Regarding the $<$$V$$>$ component of peculiar motion, the result in 
Table \ref{table:2} provides marginally significant evidence for a small
average motion counter to Galactic rotation of 
$<$$V$$>$ $=-6.2\pm3.2$ km s$^{-1}$ (1.9$\sigma$ for 10 maser)
in the middle region of the Perseus arm. 
We investigated the sensitivity of the above results to the value of the
value of the pitch angle used to define the trace of the Perseus arm.
Changing the pitch angle by $\pm2\sigma$ and recalculated the 
average peculiar motions on the interior, middle and exterior of the
arm yielded no significant changes.                                    

We now compare our observational results with basic predictions from
various models for spiral arm formation.

\subsubsection{Density wave model without shock} 
 
Linear density-wave theories which rely purely on gravity have difficulty 
explaining the radially inward motion at the interior side of the Perseus spiral arm as 
observed in the VLBI astrometric data.  This is because gas entering an overdense arm 
is accelerated gravitationally and should show radially outward motion at the interior
side of the arm as shown in Fig. 10 of \citet{2015PASJ...67...69S}.

\begin{figure*}[tbhp] 
 \begin{center} 
     \includegraphics[scale=1.0]{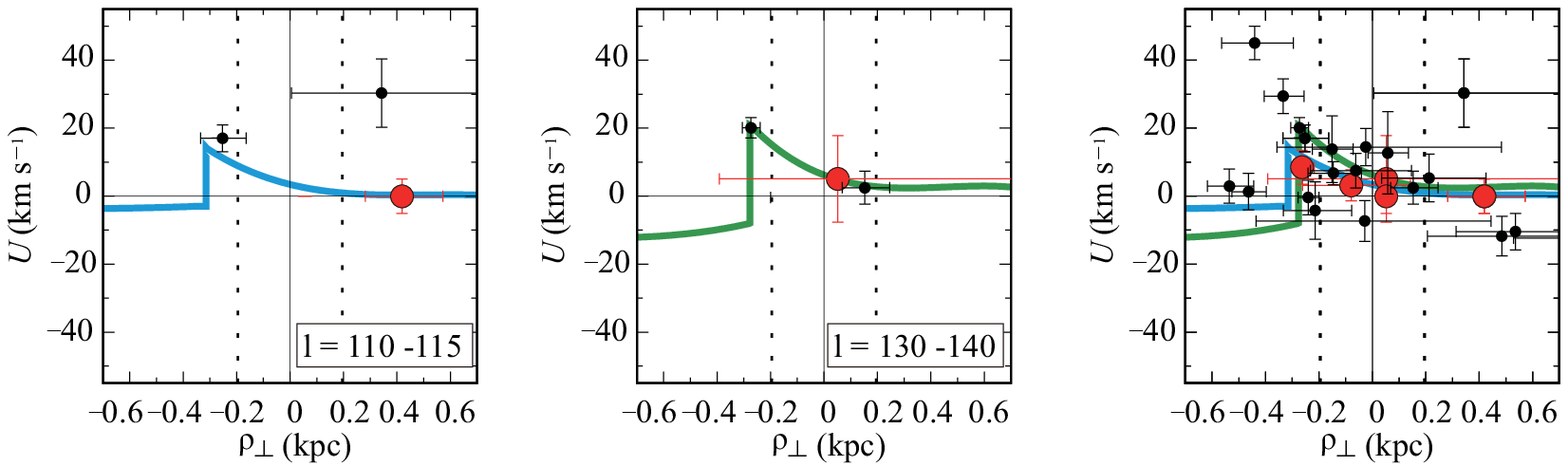} 
\end{center} 
 \caption{\textit{(Left)} Non-circular motion toward the Galactic center ($U$) is expressed as a 
function of distance perpendicular to the Perseus arm ($\rho_{\perp}$) for masers in the Galactic 
longitude range $l$ = 110$-$115 deg. Large red and small black circles show our and previous VLBI 
astrometric results, respectively.  The cyan curve represents a hydrodynamic-shock model with a 
spiral pitch angle of 12 degrees, a pattern speed of 12.5 km s$^{-1}$ kpc$^{-1}$,
a gas dispersion speed of 8 km s$^{-1}$ and a spiral potential with a 7.5\% enhancement compared 
to the axisymmetric potential (taken from Fig. 4 of \citealp{1972ApJ...173..259R}). 
\textit{(Middle)} Same as (\textit{left}) but for sources at the Galactic longitude range 
$l$ = 130$-$140 deg.   A hydrodynamic shock model shown by a green curve is 
from Fig. 3 of \citealp{1972ApJ...173..259R}. 
\textit{(Right)} Same as {\textit{left}} but for VLBI astrometric results from the entire Perseus
arm. The cyan and green curves from the other panels are superimposed.}                                 
\label{fig:4} 
\end{figure*}

\subsubsection{Density wave model with a shock} 
\begin{deluxetable*}{lcrrrrrr}
\tablecaption{Non-circular motion for Perseus, Local and Sagittarius arms.\label{table:3}}
\tablewidth{0pt}
\tablehead{
\colhead{Spiral} &Type& \multicolumn{2}{c}{Interior side} & \colhead{$\#$}& \multicolumn{2}{c}{Exterior}& \colhead{$\#$} \\
\colhead{Arm} 	&&\multicolumn{2}{c}{$-\frac{3\sigma}{2} \leqq  \rho_{\perp} \rm{(kpc)} < -\frac{\sigma}{2}$}&&\multicolumn{2}{c}{$\frac{\sigma}{2} \leqq  \rho_{\perp} \rm{(kpc)} < \frac{3\sigma}{2}$}&\\
& &\colhead{$<$$U$$>$} &\colhead{$<$$V$$>$} 	&& \colhead{$<$$U$$>$} &\colhead{$<$$V$$>$} 	\\
& &(km s$^{-1}$)	 &(km s$^{-1}$)&& (km s$^{-1}$)	&(km s$^{-1})$ 
}
\startdata
Perseus&Masers&{\bf 13.3$\pm$5.4}		&$-$2.5$\pm$2.8	&9&2.7$\pm$7.6	&3.4$\pm$2.8			&5\\
Local&Masers& 2.3$\pm$2.8		&{\bf $-$5.1$\pm$1.0}	&4&$-$4.4$\pm$4.1		&{\bf $-$6.1$\pm$2.1}		&6	\\
&OB stars&{\bf 2.7$\pm$0.9}			&0.9$\pm$1.6		&119& 5.3$\pm$3.9		&4.9$\pm$2.6	&50	\\
Sagittarius&Masers&3.1$\pm$4.6	&2.6$\pm$3.5	&3&11.1$\pm$6.8		&{\bf15.5$\pm$4.6}	&3		\\
&OB stars& 		&		&0&1.1$\pm$3.5		&{\bf 7.2$\pm$1.3}	&32	\\
\enddata
\tablecomments{Columns 1-2 indicates the spiral arm and type of observational data. Columns 3-4 show unweighted means of the non-circular
  motion components ($U$, $V$) for sources at $-\frac{3\sigma}{2} \leqq  \rho_{\perp} \rm{(kpc)} < -\frac{\sigma}{2}$, where $\sigma$ is the
  arm width and $\rho_{\perp}$ is perpendicular distance for each spiral arm (taken from \citealp{2014ApJ...783..130R}). Note that positive
  $\rho_{\perp}$ value means exterior of each arm as viewed from the Galactic center. An error in each mean shows the standard error of the
  mean. A number with the bold font indicates a statistical significance greater than 2$\sigma$. Column 5 represents the number of the sources.
  Columns 6-8 are the same as the Columns 3-5, but for sources at $\frac{\sigma}{2} \leqq  \rho_{\perp} \rm{(kpc)} < \frac{3\sigma}{2}$.}
\end{deluxetable*}

Our finding of a large positive $<$$U$$>$ value, corresponding to radially inward 
motion, for the interior side of the Perseus arm is consistent with 
density wave theories which include a shock as gas in circular Galactic
orbits encounters slower rotating spiral arms \citep{1969ApJ...158..123R, 1972ApJ...173..259R}
triggering the formation of stars. 
The hydrodynamic shock model of \citet{1972ApJ...173..259R} with a pitch angle, $\psi$, of $12^\circ$,
a pattern speed, $\Omega_{{\rm p}}$, of 12.5 km~s$^{-1}$~kpc$^{-1}$, a gaseous dispersion speed, $a$, of
8 km~s$^{-1}$, and a spiral potential with an enhancement, $F$, of 7.5\% compared to
the axisymmetric potential, predicts a velocity jump in front of the gravitational 
potential minimum as shown by Fig. 1 of \citet{1972ApJ...173..259R}.

If we assume the full jump velocity is in the line-of-sight direction, the line-of-sight 
vector of \citet{1972ApJ...173..259R} model can be decomposed into $U$ (and $V$) jumps by subtracting the
rotation curve model of \citet{1972ApJ...173..259R}. This assumption is reasonable for the Perseus arm at
the Galactic longitude range $l$ = 110$-$140 (deg) since the non-circular motion vectors in Fig. \ref{fig:2}
are aligned in the line-of-sight direction at the section. The $U$ jumps with amplitudes of 20$-$30 km s$^{-1}$,
shown by cyan and green curves in Fig. \ref{fig:4}, indicate large positive $<$$U$$>$ values for the interior side of the Perseus arm and no significant $<$$U$$>$ values for the
exterior of the arm, which are consistent with the observational results. 
We also investigated the shock model of \citet{1972ApJ...173..259R} with $\psi$ = 8$^{\circ}$ and 
$F$ = 5\% (taken from his Figures 7 and 8) and found similar characteristics, but with 
shock velocities decreased to 10$-$20 km s$^{-1}$ and the shock locations shifted by $<$ 150 pc. 

\subsubsection{Dynamic spiral arm formation} 
 
We now compare the observational results with a dynamic spiral-arm model
proposed by \citet{2018ApJ...853L..23B}. The dynamic spiral-arm model is a barred spiral galaxy
generated from $N-$body$/$hydrodynamics simulations, and amplitudes, pitch angles,
and pattern speeds of spiral arms change within a few hundred million years. \citet{2018ApJ...853L..23B}
picked spiral arms in growth and disruption phases, respectively, from the model. 
The growth phase has negative $<$$U$$>$ value for the interior side of an arm and
thus is inconsistent with our observational results.  While the disruption phase has
positive $<$$U$$>$ for the interior side of the arm, it has negative values for the exterior 
of the arm, and thus also does not agree with our observational results.
  
\subsection{Universality of non-circular arm motions}

We confirm that the shock model of \citet{1972ApJ...173..259R} can explain the 
observed radially inward motions in the interior side of the Perseus arm. 
In order to investigate the universality of these motions, we examine the 
non-circular motions for other spiral arms using the same procedure
applied to the Perseus arm and the VLBI astrometric results compiled
in \citet{2014ApJ...783..130R}.  We also examine $Gaia$ DR2 results for OB-type stars 
taken from Fig. 2(a) of \citet{2018A&A...616L..15X}.
Table \ref{table:3} displays the non-circular motion components ($<$$U$$>$, $<$$V$$>$) for the 
interior and exterior of the spiral arms.   While there are some statistically significant
average motions, no clear trend is evident for these spiral arms, suggesting a more complex 
picture than expected from the basic models discussed above.

\section{Summary}
We presented parallaxes and proper motions for five 6.7-GHz methanol masers associated with 
HMSFRs in the outer portion of the Perseus spiral arm as part of the BeSSeL Survey of the Galaxy 
(see Figure \ref{fig:1} and Table \ref{table:1}). Combining these new and previous VLBI results, we determined a spiral-arm pitch angle 
of 9.2 $\pm$ 1.5 deg  and an arm width of 0.39 kpc (see Section \ref{sec:5.1}). 

We divided the sources into interior, middle and exterior regions of the Perseus arm 
and averaged the non-circular motion components ($<U>$, $<V>$, $<W>$) for each region. For nine sources 
in the interior of the arm, we found a radially inward motion of $<U>$ = 13.3 $\pm$ 5.4 km s$^{-1}$; 
for 10 sources in the middle of the arm, we obtained a marginal detection of motion slower than Galactic rotation 
of $<V>$ = $-$6.2 $\pm$ 3.2 km s$^{-1}$; and for 5 sources in the exterior of the arm, we found no statistically
significant non-circular motion (see Section \ref{sec:5.2} and Table \ref{table:2}). These characteristics are
consistent with predictions of models for spiral arm formation that involve gas entering an arm to be shocked 
and then forming stars as shown by Fig. \ref{fig:4}.

We performed a similar analysis on previous VLBI astrometric data, as well
as on $Gaia$ DR2 results for OB-type stars, for stars in other spiral arms.  While some statistically 
significant non-circular motions are found in other arms, no clear pattern among arms was found
(see Table \ref{table:3}). This suggests a more complex picture than expected from basic spiral-arm models.


\acknowledgments 
 We acknowledge anonymous referee for valuable comments, which improved the manuscript.

\vspace{5mm} 
 
\facility{VLBA}. 
  
 

\bibliographystyle{apj}
\bibliography{apj-jour,reference}

\appendix 
 
\section{Appendix} 
 
\begin{figure*}[tbhp] 
 \begin{center} 
     \includegraphics[scale=1.15]{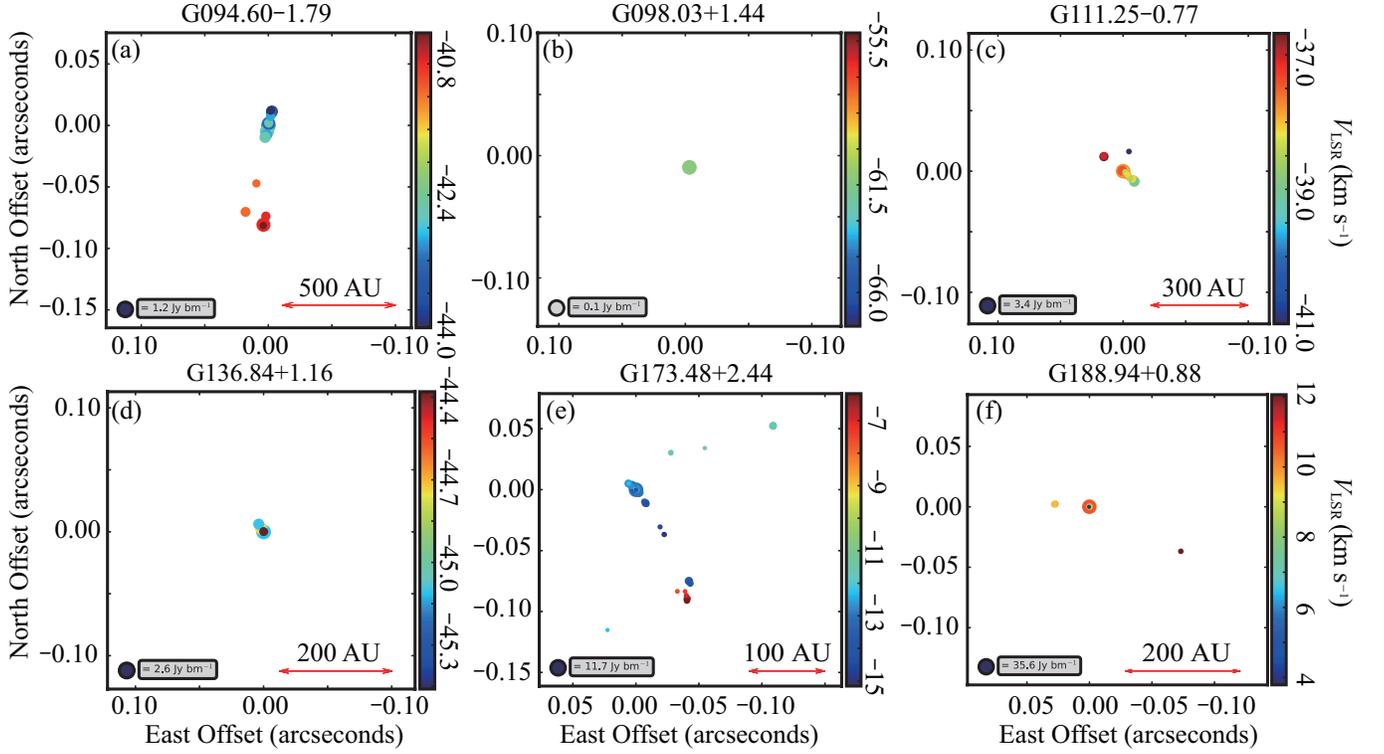} 
\end{center} 
    \caption{Maser spot distributions for {\bf (a)} G094.60$-$1.79, {\bf(b)} G098.03+1.44,
{\bf(c)} G111.25$-$0.76, {\bf(d)} G136.84+1.16, {\bf(e)} G173.48+2.44 and {\bf(f)} G188.94+0.88. 
 The distributions were made using 1st epoch data of individual sources. 
The origin of coordinates for each map is described in Table \ref{table:4}.
The horizontal red arrow in each map, except for G098.03+1.44, shows
an absolute spatial scale converted at a source distance (see Table \ref{table:1}). 
Color bar indicates the local standard of
rest (LSR) velocity. The size of a maser spot is proportional
to (Jy/beam).}    
     \label{fig:5} 
\end{figure*}

Here, we show supplemental materials to further document observations 
(Table \ref{table:4}) and detailed maser maps for (Fig. \ref{fig:5}).                               
  
\begin{table*}[htbp] 
\caption{Observational Information.} 
\label{table:4} 
\small 
\begin{tabular}{llllllll} 
\tableline 
\tableline 
Project &Source&R.A. &Decl. &Epoch 1	  &Epoch 2	&Epoch 3	    &Epoch
4        		\\                                                           
	    &          &(hh:mm:ss) &(dd:mm:ss) &(in 2012)&(in 2013)&(in
2013) &(in 2013)\\                                                      
\tableline 
 
BR149S &G136.84+1.16	&02:49:33.609  	&+60:48:27.92&Dec 08	&May 19
			&June 24			&Nov 24		\\                                              
 &J0244+6228	&02:44:57.6966  	&+62:28:06.517&\ \ \ \ 	& 	      &	&		\\       
 &J0248+6214	&02:48:58.8920  	&+62:14:09.678&\ \ \ \ &      &	&		\\       
 &J0306+6243	&03:06:42.6595  	&+62:43:02.024&\ \ \ \ 			&      &	&	\\       
 
\tableline 
BR149T &G173.48+2.44	&05:39:13.066  &+35:45:51.28&Sep 22	&March 09
			&April 05			&Sep 13	\\                                               
  &J0530+3723	&05:30:12.5493  	&+37:23:32.620&\ \ \ \ 
&	      &					&	\\       
  &J0539+3308	&05:39:09.6722  	&+33:08:15.496&\ \ \ \ 
		&			 	      &	&	\\       
 &J0541+3301	&05:41:49.4359  	&+33:01:31.890&\ \ \ \ 
				&	      &&	\\       
  &J0552+3754	&05:52:17.9369  	&+37:54:25.281&\ \ \ \ 
				&	      &	&	\\  \\
  &G188.94+0.88	&06:08:53.341		&+21:38:29.08&\ \ \
			&      &	&	\\     
  &J0603+2159	&05:30:12.5493  	&+37:23:32.620&\ \ \ \ 
	&      &	&	\\       
  &J0607+2129	&06:07:59.5657  	&+21:29:43.720&\ \ \ \ 
				&      &	&	\\       
  &J0607+2218	&06:07:17.4360  	&+22:18:19.080&\ \ \ \ 
				&      &	&	\\       
  &J0608+2229	&06:08:34.3109  	&+22:29:42.981&\ \ \ \ 
				&	      &	&	\\       
\tableline 
BR149U &G094.60$-$1.79	&21:39:58.258 			&+50:14:21.02&Dec 03	&May
12 			&June 06			&Nov 23	\\                                             
  &J2137+5101	&21:37:00.9862  	&+51:01:36.129&\ \ \ \ 
				&      &	&	\\       
  &J2145+5147	&21:45:07.6666  	&+51:47:02.243&\ \ \ \ 
	&      &	&	\\       
  &J2150+5103	&21:50:14.2662  	&+51:03:32.264&\ \ \ \ 
	&      &	&	\\       
  &(J2139+5300)	&21:39:53.6244  	&+53:00:16.599&\ \ \ \
				&      &	&	\\    \\
 
     &(G098.03+1.44)		&21:43:01.431 			&+54:56:17.72&\
				&      &	&	\\ 
  &J2123+5452	&21:23:46.8349  	&+54:52:43.488&\ \ \ \ 
				&      &	&	\\       
 &(J2139+5300)	&21:39:53.6244  	&+53:00:16.599&\ \ \ \
	&      &	&	\\       
  &J2139+5540	&21:39:32.6175  	&+55:40:31.771&\ \ \ \ 
	&      &	&	\\       
  &J2145+5147	&21:45:07.6666  	&+51:47:02.243&\ \ \ \ 
	&	      &	&	\\   \\
 
     &G111.25$-$0.76	&23:16:10.327 	&+59:55:28.66&\ \
	&      &	&	\\   
  &J2339+6010	&23:39:21.1252  	&+60:10:11.849&	&	 &	&	\\       
  &J2254+6209	&22:54:25.2926  	&+62:09:38.723&\ \ \ \ 
	&      &	&	\\       
  &J2301+5706	&23:01:26.6271  	&+57:06:25.499&\ \ \ \ 
	&      &	&	\\       
  &(J2314+5813)	&23:14:19.0833  	&+58:13:47.647&\ \ \ \
				&	      &	&	\\       
 
\tableline 
\end{tabular} 
\tablecomments{Column 1 shows project name. Column 2 lists an observed
6.7 GHz CH$_{3}$OH maser source (as denoted by ``G") and background
QSOs (as denoted by ``J"). Parenthesis indicates
an extended source, which was removed from the parallax determination.
Columns 3-4 represent equatorial coordinates for the source in (J2000). Columns
5-8 show dates of observations.}                                        
\end{table*}

\end{document}